\documentclass[aip,apl,reprint,graphicx]{revtex4-1} 
\usepackage{graphicx}
\usepackage{amsmath,amsfonts}
\usepackage{natbib}
\begin{document}

\title{Tuning the spontaneous exchange bias effect with Ba to Sr partial substitution in La$_{1.5}$(Sr$_{0.5-x}$Ba$_{x}$)CoMnO$_{6}$}

\author{M. Boldrin}
\affiliation{Instituto de F\'{\i}sica, Universidade Federal de Goi\'{a}s, 74001-970, Goi\^{a}nia, GO, Brazil}

\author{A. G. Silva}
\affiliation{Instituto de F\'{\i}sica, Universidade Federal de Goi\'{a}s, 74001-970, Goi\^{a}nia, GO, Brazil}

\author{L. T. Coutrim}
\affiliation{Instituto de F\'{\i}sica, Universidade Federal de Goi\'{a}s, 74001-970, Goi\^{a}nia, GO, Brazil}

\author{J. R. Jesus}
\affiliation{Centro Brasileiro de Pesquisas F\'{\i}sicas, 22290-180, Rio de Janeiro, RJ, Brazil}

\author{C. Macchiutti}
\affiliation{Centro Brasileiro de Pesquisas F\'{\i}sicas, 22290-180, Rio de Janeiro, RJ, Brazil}

\author{E. M. Bittar}
\affiliation{Centro Brasileiro de Pesquisas F\'{\i}sicas, 22290-180, Rio de Janeiro, RJ, Brazil}

\author{L. Bufai\c{c}al}
\email{lbufaical@ufg.br}
\affiliation{Instituto de F\'{\i}sica, Universidade Federal de Goi\'{a}s, 74001-970, Goi\^{a}nia, GO, Brazil}

\date{\today}

\begin{abstract}

The spontaneous exchange bias (SEB) effect is a remarkable phenomenon recently observed in some reentrant spin-glass materials. Here we investigate the SEB in La$_{1.5}$(Sr$_{0.5-x}$Ba$_{x}$)CoMnO$_{6}$ double-perovskites, a system with multifarious magnetic phases for which a notable increase in the exchange bias field is observed for intermediate Sr/Ba concentrations. The Ba to Sr substitution leads to the enhancement of the crystal lattice, which is accompanied by the raise of both the effective magnetic moment ($\mu_{eff}$) and the antiferromagnetic (AFM) transition temperature that is observed below the ferromagnetic ordering. Such increases are likely related to the increased fraction of Co$^{3+}$ in the high spin configuration, leading to the enhancement of Co$^{3+}$--O--Mn$^{4+}$ AFM phase and to the reduction in the uncompensation of the AFM coupling between Co and Mn. The combined effect of the increased $\mu_{eff}$ and AFM phase plausible explains the changes in the SEB effect.

\end{abstract}


\maketitle

Conventionally, the exchange bias (EB) effect is manifested as a shift in hysteresis loop measurements [$M(H)$] carried after cooling the system in the presence of an external magnetic field ($H$) \cite{Nogues}. Due to its potential application in ultra-dense magnetic recording heads, magnetoresistive devices and spin valves, this phenomena has been extensively investigated in the last decades. For many years it was thought that the cooling $H$ was a necessary condition to set the unidirectional anisotropy (UA) by breaking the symmetry at the magnetic phase interfaces. Recently, however, it was observed that some materials exhibit such asymmetry even after being cooled in zero $H$ \cite{Wang,Nayak,Maity}. These are the so-called spontaneous EB (SEB) materials, for which the EB effect observed at low $T$ seems to be related to the presence of spin glass (SG)-like phases \cite{PRB2018,New_SEB}.

There is an effort of the academic community to find new materials exhibiting robust SEB at higher $T$. In this sense, the synthesis and investigation of new materials exhibiting such effect is mandatory. Double-perovskite (DP) compounds are prospective candidates since their intrinsic structural and magnetic inhomogeneities usually leads to magnetic phase competition and frustration, key ingredients to achieve glassy magnetic behavior \cite{Sami,Mydosh}. Among the DP materials, La$_{1.5}$Sr$_{0.5}$CoMnO$_{6}$ stands out by presenting a giant SEB effect \cite{Murthy}. The Ba substitution also leads to a very large SEB in La$_{1.5}$Ba$_{0.5}$CoMnO$_{6}$ \cite{Ba2020}, whilst for La$_{1.5}$Ca$_{0.5}$CoMnO$_{6}$ the effect is very subtle \cite{Ca2017}. A detailed investigation of these materials' magnetic and electronic properties by means of X-ray absorption spectroscopy (XAS) and X-ray magnetic circular dichroism (XMCD) has shown that the UA in these compounds results from uncompensated antiferromagnetic (AFM) coupling between Co and Mn ions. The distinct magnitude of the SEB effect observed in each compound is related to the proportion of Co$^{3+}$ high spin (HS) and non-magnetic low spin (LS) configuration \cite{PRB2019}.

The SEB effect in La$_{1.5}$A$_{0.5}$CoMnO$_{6}$ (A = Ba, Sr, Ca) is directly related to the Co$^{3+}$ spin state, which in turn results from a delicate balance between the crystal field and the interatomic exchange energies, being thus very sensitive to structural changes \cite{Raveau}. In this work we investigate the changes in the SEB when Sr is partially replaced by Ba in the La$_{1.5}$(Sr$_{0.5-x}$Ba$_{x}$)CoMnO$_{6}$ series. A detailed investigation of the structural, electronic, and magnetic properties of the system was carried by means of X-ray powder diffraction (XRD) and magnetometry. Our results show a pronounced enhancement of the SEB effect for intermediate Sr/Ba concentrations. This can be causally linked to the changes in the balance between the crystal field and the intra-orbital Coulomb repulsion acting on the Co ions, affecting its spin state and consequently its coupling with the Mn ions.

Polycrystalline samples of La$_{1.5}$(Sr$_{0.5-x}$Ba$_{x}$)CoMnO$_{6}$ ($x$ = 0, 0.2, 0.25, 0.3, 0.4 and 0.5) were prepared by conventional solid state reaction, as described in the Supplementary Material (SM). Attempts to produce 0 $<x<$ 0.2 samples by using the same synthesis route were not successful. Laboratory XRD data were collected using a Bruker $D$8 $Discover$ diffractometer operating with Cu K$_{\alpha}$ radiation, while high resolution synchrotron XRD (SXRD) patterns were recorded at the XPD beamline of the Brazilian Synchrotron Light Laboratory (LNLS), using a reflection geometry. The data were obtained at room $T$ by one-dimensional Mythen-1K detector (Dectris), using $\lambda$ = 1.9074 $\textrm{\AA}$. Rietveld refinements were performed using the GSAS+EXPGUI suite\cite{GSAS}. Magnetic measurements were carried out in both zero field cooled (ZFC) and field cooled (FC) modes, using a Quantum Design PPMS-VSM magnetometer.

\begin{figure}
\includegraphics[width=0.46 \textwidth]{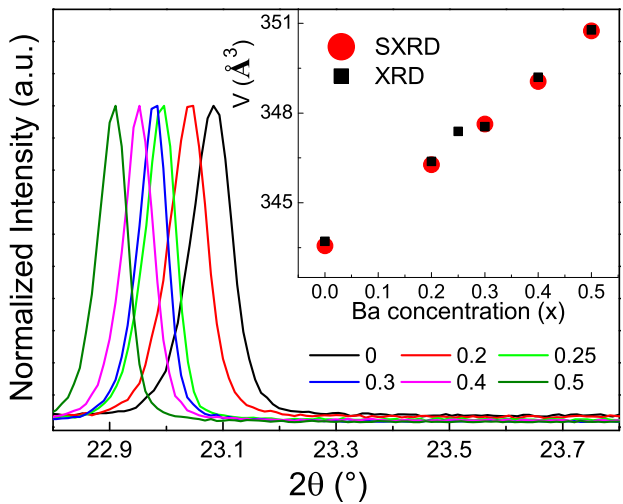}
\caption{Magnified view of the ($10\bar{2}$) Bragg reflection in the normalized XRD patterns of La$_{1.5}$(Sr$_{0.5-x}$Ba$_{x}$)CoMnO$_{6}$. The inset shows the unit cell volume as a function of Ba concentration ($x$), obtained from XRD (squares) and SXRD (circles).}
\label{Fig_XRD}
\end{figure}

The XRD patterns revealed that all samples belong to the rhombohedral $R\bar{3}c$ space group, as expected since both $x$ = 0 and 0.5 end members present the same crystal symmetry \cite{Murthy,Ba2020}. Fig. \ref{Fig_XRD} shows a magnified view of the XRD data near the ($10\bar{2}$) Bragg reflection, evidencing a systematic shift of the peak toward smaller angles as the Ba concentration increases. This is a manifestation of the lattice expansion, reflecting the fact that the Ba$^{2+}$ ionic radius is larger than Sr$^{2+}$ \cite{Shannon}. Such lattice expansion with increasing Ba content is also manifested in the increase of the unit cell volume, obtained from the Rietveld refinements of both XRD and SXRD data (inset of Fig. \ref{Fig_XRD}). Main results obtained from the XRD Rietveld refinements are displayed in Table \ref{T1}. Results from SXRD patterns are in Table S1 of SM. The crystal symmetry of DP compounds is directly related to the hybridization between the $d$ orbitals, via oxygen $p$ orbitals, being thus strongly correlated to their electronic and magnetic properties \cite{Sami,Woodward}. Accordingly, the structural changes here observed will remarkably affect the materials magnetic properties, in special those associated to the Co spin state.

\begin{table}
\caption{Main results obtained from the Rietveld refinements of the XRD data for the La$_{1.5}$(Sr$_{0.5-x}$Ba$_{x}$)CoMnO$_{6}$ series.}
\label{T1}
\begin{tabular}{cccccc}
\hline \hline
Sample ($x$) & $a$ (\AA) & $c$ (\AA) & $V$ (\AA$^{3}$) & $R_{wp}$ (\%) & $R_{p}$ (\%) \\
\hline
0 & 5.4733(1) & 13.2575(2) & 343.95(1) & 10.3 & 7.3 \\
0.2 & 5.4852(1) & 13.2937(1) & 346.38(1) & 10.8 & 8.0 \\
0.25 & 5.4886(1) & 13.3157(1) & 347.39(1) & 8.8 & 6.4 \\
0.3 & 5.4885(1) & 13.3221(1) & 347.54(1) & 10.1 & 7.6 \\
0.4 & 5.4952(1) & 13.3531(1) & 349.20(1) & 11.1 & 8.2 \\
0.5 & 5.5025(1) & 13.3777(1) & 350.78(1) & 12.1 & 8.7 \\
\hline \hline
\end{tabular}
\end{table}

\begin{figure}
\includegraphics[width=0.46 \textwidth]{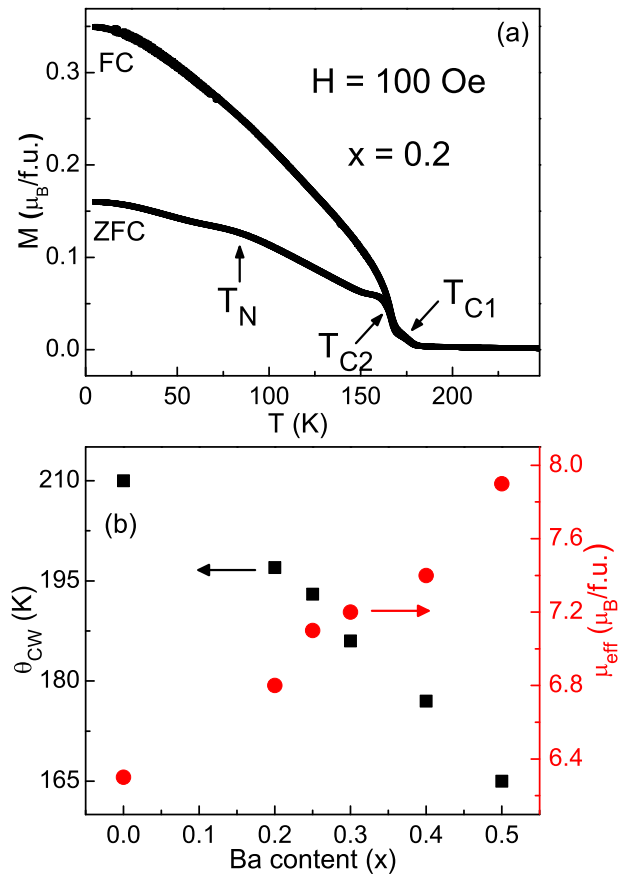}
\caption{(a) ZFC and FC $M(T)$ curves measured at $H$ = 100 Oe for $x$ = 0.2. (b) $\theta_{CW}$ (square) and $\mu_{eff}$ (circle) as a function of $x$ for La$_{1.5}$(Sr$_{0.5-x}$Ba$_{x}$)CoMnO$_{6}$.}
\label{Fig_MxT}
\end{figure}

Figure \ref{Fig_MxT}(a) shows the ZFC and FC magnetization as a function of $T$ [$M(T)$] curves of $x$ = 0.2 compound, chosen here as a representative sample for the series. The FC curve shows a ferromagnetic (FM)-like character, while the ZFC one highlights three anomalies associated to distinct magnetic couplings. The same overall behavior was observed for all samples (see Fig. S2 in SM). Previous reports of magnetization, muon spin rotation and relaxation, XAS and XMCD measurements for the $x$ = 0 and 0.5 compounds have shown a mixed valence state for Co (Co$^{2+}$/Co$^{3+}$), while for Mn it was found a nearly tetravalent state, although some small amount of Mn$^{3+}$ was detected \cite{Murthy,Ba2020,PRB2019}. In these studies the first transition temperature, $T_{C1}$, was ascribed to the Co$^{2+}$--O--Mn$^{4+}$ FM coupling, the second one, $T_{C2}$, to the Co$^{3+}$--O--Mn$^{3+}$ FM coupling, while the third anomaly observed at lower $T$ is most likely related to the Co$^{3+}$--O--Mn$^{4+}$ AFM coupling. It is natural to consider the same scenario for the other Ba/Sr concentrations here investigated. The transition temperatures obtained from the first derivative of the ZFC curves (see Figs. S4 and S5 in SM) are displayed in Table \ref{T2}, where it is observed a systematic increase of $T_N$ with the Ba content. Previous works have also found SG-like behavior for both $x$ = 0 and 0.5 end members at low temperatures \cite{Murthy,PRB2019}. Since all samples here investigated present the same overall magnetic behavior, SG-like features are also expected for the intermediate Ba/Sr compounds.

From the fit of the magnetic susceptibility curves at the paramagnetic region, with the Curie-Weiss (CW) law (see Fig. S3 in SM), we obtain the CW temperature ($\theta_{CW}$) and the effective magnetic moment ($\mu_{eff}$). As it can be seen in Table \ref{T2}, $\theta_{CW}$ is positive for all investigated samples, giving further evidence of the prominent FM coupling. Its systematically decrease with increasing the Ba content ($x$) is in accordance with the increase of the AFM transition temperature ($T_N$).

\begin{table*}
\caption{Main results obtained from the $M(T)$ and $M(H)$ curves for the La$_{1.5}$(Sr$_{0.5-x}$Ba$_{x}$)CoMnO$_{6}$ series.}
\label{T2}
\begin{tabular}{c|cccccc|ccc}
\hline \hline
Sample ($x$) &  $T_{C1}$ (K)  & $T_{C2}$ (K)  & $T_N$ (K)  &  $\theta_{CW}$ (K)  &  $\mu_{eff}$ ($\mu_B$/f.u.)  &  C ($K\cdot\mu_B$/f.u.)  &  $H_{EB}$ (Oe)  &  $H_C$ (Oe) &  $M_S$ ($\mu_B$/f.u.) \\
\hline
0 & 185 & 178 & 77 & 210 & 6.3 & 1323 & 5112 & 10968 & 2.01 \\
0.2 & 175 & 165 & 84 & 197 & 6.8 & 1340 & 7119 & 6128 & 2.44  \\
0.25 & 181 & 173 & 88 & 193 & 7.1 & 1370 & 9343 & 4629 & 2.26 \\
0.3 & 182 & 171 & 90 & 186 & 7.2 & 1339 & 8528 & 4699 & 2.11 \\
0.4 & 187 & 176 & 106 & 177 & 7.4 & 1310 & 7657 & 4022 & 1.96 \\
0.5 & 181 & 165 & 110 & 165 & 7.9 & 1304 &  4261 & 5792 & 2.41 \\
\hline \hline
\end{tabular}
\end{table*}

The $\mu_{eff}$ systematically increases with $x$, plausibly explaining the increase in $T_N$. As aforementioned, previous studies of Co- and Mn-based DPs have revealed the presence of Co$^{2+}$/Co$^{3+}$ mixed valence, with the amounts of LS/HS Co$^{3+}$ playing an important role on the SEB effect \cite{PRB2019}. Such investigations have also shown that changes in the A site of the perovskite structure act mainly in the Co electronic configuration, while the Mn formal valence remains nearly close to 4+. Thus, the enhancement of $\mu_{eff}$ is most likely related to the increase in the amount of Co$^{3+}$ in high HS configuration present in the system. This is endorsed by the decrease of $\theta_{CW}$ accompanied by the increase of $T_N$, which are likely related to the (HS)Co$^{3+}$--O--Mn$^{4+}$ AFM coupling \cite{PRB2019}. These changes will cause a direct impact in the SEB effect of the La$_{1.5}$(Sr$_{0.5-x}$Ba$_{x}$)CoMnO$_{6}$ series as detailed below.

\begin{figure}
\includegraphics[width=0.46 \textwidth]{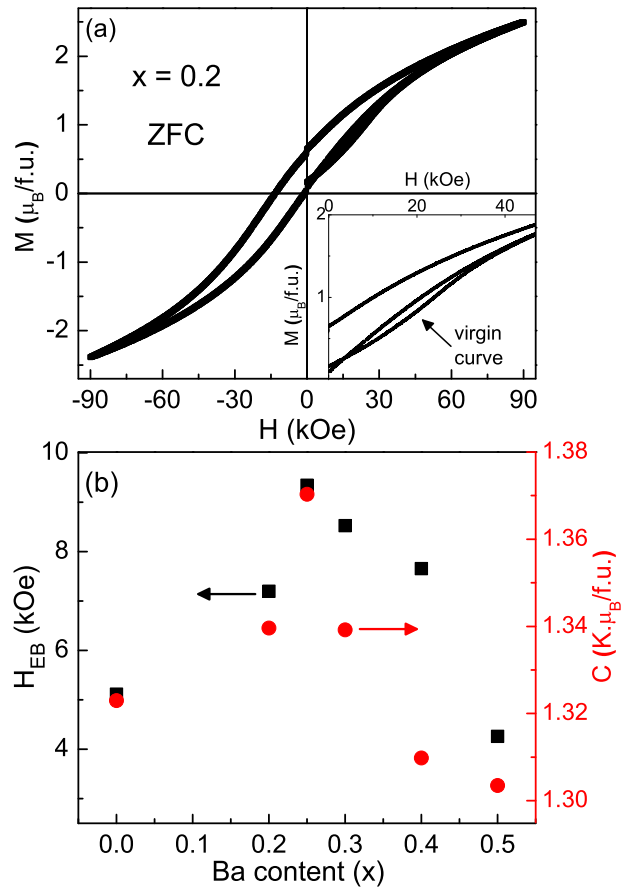}
\caption{(a) ZFC $M(H)$ loop for the $x$ = 0.2 sample, measured at 5 K. Inset shows a magnified view of the initial magnetization curve (virgin curve). (b) $H_{EB}$ (square) and $C = \mu_{eff}\cdot\theta_{CW}$ (circle) as a function of $x$ (see text).}
\label{Fig_MxH}
\end{figure}

Fig. \ref{Fig_MxH}(a) shows the ZFC $M(H)$ curve of the $x$ = 0.2 sample, measured at 5 K. It is a closed loop and its shape results from a combined contribution of AFM and FM phases. The shift towards the left along the $H$ axis characterizes the SEB effect. Inset shows that the virgin magnetization curve falls outside the hysteresis loop up to a certain field. This is a common feature of several SEB DP compounds, being related to field induced spin transition \cite{Murthy,Ba2020,Ca2017,Xie}. The same overall behavior was observed for the $M(H)$ curves of all investigated samples, as shown in Fig. S6 of SM. The shift along the $H$ axis is a measure of the EB field, herein defined as $H_{EB}=|H_1+H_2|/2$, where $H_1$ and $H_2$ are respectively the left and right coercive fields. Fig. \ref{Fig_MxH}(b) shows a pronounced increase of the SEB for intermediate contents of Ba and Sr, being maximum for the $x$ = 0.25 sample.

The evolution of the SEB effect in La$_{1.5}$(Sr$_{0.5-x}$Ba$_{x}$)CoMnO$_{6}$ can be understood within the framework of strong correlation between the structural and electronic properties observed for this system. The partial replacement of Sr by Ba acts to increase the crystal lattice, leading to the decrease of the crystal field splitting energy and thus favoring the HS configuration. This is manifested in the increase of $\mu_{eff}$ and in the decrease of $\theta_{CW}$, which are related to the increased strength of the Co$^{3+}$--O--Mn$^{4+}$ AFM coupling. On the other hand, such enhancement in the Co magnetic moment reduces the uncompensation of its coupling with Mn, which would tend to reduce the EB field \cite{PRB2019}. 

The lattice expansion is closely related to the increase in the portion of HS Co$^{3+}$, which in turn brings two main outcomes: (i) the increase in the strength of the AFM coupling and (ii) the decrease in the uncompensated coupling between the transition-metal (TM) ions. These opposite effects are causally related to the changes in the SEB effect. In order to quantify this argument, we define here a coupling parameter $C = \mu_{eff}\cdot\theta_{CW}$. Fig. \ref{Fig_MxH}(b) shows the evolution of $C$ as a function of $x$. As can be noticed, it presents the same behavior of the $H_{EB}$ curve. One can therefore conclude that the increase in the portion of HS Co$^{3+}$ with $x$, manifested in the enhancement of $\mu_{eff}$, leads to the rise of the AFM phase, which is necessary for the observation of the SEB effect. In addition, it also leads to the a decrease in the uncompensation of the coupling between the Co and Mn superstructures, which acts to decrease the SEB. The intermediate concentration of Sr/Ba is the region of the series where this joint effect is maxima, resulting in the pronounced increase of $H_{EB}$.

J. Krishna Murthy \textit{et al.} have found a maxima in the SEB of La$_{2-x}$Sr$_x$CoMnO$_6$ (0$\leq$x$\leq$1) for intermediate $x$ concentrations \cite{Murthy2}. Such effect was explained in terms of the antisite disorder (ASD) at the Co/Mn site, which destroys the long-range magnetic order. The UA is set by a field-induced metamagnetic phase transformation from canted AFM to FM during the initial magnetization process of the magnetic hysteresis cycle, and the changes in SEB of La$_{2-x}$Sr$_x$CoMnO$_6$ are causally linked to the ASD via the ``saturated" magnetization in the $M(H)$ curves (the magnetization at $H$ = 70 kOe, $M_S$). In the case of La$_{1.5}$(Sr$_{0.5-x}$Ba$_x$)CoMnO$_6$, although it is not observed a direct relation between $H_{EB}$ and the $M_S$ values [magnetization at $H$ = 90 kOe in the $M(H)$ curves, see Table \ref{T2}], the scenario of phase separation is also plausible. The increment of Ba leads to the increase of the AFM phase due to the increased portion of HS Co$^{3+}$, but it also affects the FM coupling by increasing the distance between the TM ions. Since the EB is an interface effect, it will depend on the balance between the magnetic phases, being maxima in the intermediate Sr/Ba region of the series.

In summary, we thoroughly investigated the structural, electronic, and magnetic properties of the La$_{1.5}$(Sr$_{0.5-x}$Ba$_{x}$)CoMnO$_{6}$ system by means of XRD and magnetometry measurements. The lattice expansion with increasing $x$ is closely related to the increase of $\mu_{eff}$ and $T_N$. The ZFC $M(H)$ loops showed a pronounced increase of the giant SEB for intermediate concentrations of Sr/Ba, where the combined effects of the enhanced AFM phase and the decreased uncompensation in the magnetic coupling between Co and Mn, both resulting from the increased portion of HS Co$^{3+}$, result in a maximum SEB effect for the $x$ = 0.25 compound.

\section*{Supplementary Material}

See supplementary material for details of the crystals growth and some structural and magnetic properties of the investigated samples.

\begin{acknowledgments}
This work was supported by Conselho Nacional de Desenvolvimento Cient\'{i}fico e Tecnol\'{o}gico (CNPq) (Grant No. 400134/2016-0), Funda\c{c}\~{a}o Carlos Chagas Filho de Amparo \`{a} Pesquisa do Estado do Rio de Janeiro (FAPERJ) (Grant No. E-26/202.798/2019), Funda\c{c}\~{a}o de Amparo \`{a} Pesquisa do Estado de Goi\'{a}s (FAPEG) and Coordena\c{c}\~{a}o de Aperfei\c{c}oamento de Pessoal de N\'{i}vel Superior (CAPES).
\end{acknowledgments}

\section*{Data Availability Statement}
The data that support the findings of this study are available from the corresponding author upon reasonable request.


\begin{thebibliography}{99}

\bibitem{Nogues} J. Nogu\'{e}s and I. K. Schuller, J. Magn. Magn. Mater. \textbf{192} 203 (1999).

\bibitem{Wang} B. M. Wang, Y. Liu, P. Ren, B. Xia, K. B. Ruan2 J. B. Yi, J. Ding, X. G. Li, and L. Wang, Phys. Rev. Lett. \textbf{106}, 0077203 (2011).

\bibitem{Maity} T. Maity, S. Goswami, D. Bhattacharya, and S. Roy, Phys. Rev. Lett. \textbf{110}, 107201 (2013).

\bibitem{Nayak} A. K. Nayak, M. Nicklas, S. Chadov, C. Shekhar, Y. Skourski, J. Winterlik, and C. Felser, Phys. Rev. Lett. \textbf{110}, 127204 (2013).

\bibitem{PRB2018} L. T. Coutrim, E. M. Bittar, F. Garcia, and L. Bufai\c{c}al, Phys. Rev. B \textbf{98}, 064426 (2018).

\bibitem{New_SEB} L. Bufai\c{c}al, L. T. Coutrim, E. M. Bittar, and F. Garcia, J. Magn. Magn. Mater. \textbf{512} 167048 (2020).

\bibitem{Sami} S. Vasala and M. Karppinen, Prog. Solid State Chem. \textbf{43}, 1 (2015).

\bibitem{Mydosh} J. A. Mydosh, \textit{Spin Glasses: An Experimental Introduction} (Taylor \& Francis, London, 1993).

\bibitem{Murthy} J. Krishna Murthy and A. Venimadhav, Appl. Phys. Lett. \textbf{103}, 25410 (2013).

\bibitem{Ba2020} M. Boldrin, L. T. Coutrim and L. Bufai\c{c}al, Braz. J. Phys. \textbf{50}, 711 (2020).

\bibitem{Ca2017} L. Bufai\c{c}al, R. Finkler, L. T. Coutrim, P. G. Pagliuso, C. Grossi, F. Stavale, E. Baggio-Saitovitch, and E. M. Bittar, J. Magn. Magn. Mater. \textbf{433}, 271 (2017).

\bibitem{PRB2019} L. T. Coutrim, D. Rigitano, C. Macchiutti, T. J. A. Mori, R. Lora-Serrano, E. Granado, E. Sadrollahi, F. J. Litterst, M. B. Fontes, E. Baggio-Saitovitch, E. M. Bittar, and L. Bufai\c{c}al, Phys. Rev. B \textbf{100}, 054428 (2019).

\bibitem{Raveau} B. Raveau and Md. Motin Seikh, \textit{Cobalt Oxides: From Crystal Chemistry to Physics} (Wiley-VCH, Weinheim, 2012).

\bibitem{GSAS} A. C. Larson and R. B. Von Dreele, Los Alamos National Laboratory Report No. LAUR 86-748, 2000; B. H. Toby, J. Appl. Crystallogr. \textbf{34}, 210 (2001).

\bibitem{Shannon} R. D. Shannon, Acta Crystallographica A32 (1976) 751.

\bibitem{Woodward} C. J. Howard, B. J. Kennedy and P. M. Woodward, Acta Crystalogr. B \textbf{59}, 463 (2003).

\bibitem{Xie} L. Xie and H. G. Zhang, Curr. Appl. Phys. \textbf{18} 261 (2018).

\bibitem{Murthy2} J. Krishna Murthy, K. D. Chandrasekhar, H. C. Wu, H. D. Yang, J. Y. Lin, and A. Venimadhav, J. Phys.: Condens. Matter \textbf{28}, 086003 (2016).

\end{thebibliography}
\end{document}